\documentclass[conference]{IEEEtran}
\IEEEoverridecommandlockouts
\usepackage{cite}
\usepackage{amsmath,amssymb,amsfonts}
\usepackage{algorithmic}
\usepackage{graphicx}
\usepackage{textcomp}
\usepackage{xcolor}
\usepackage{hyperref}
\def\BibTeX{{\rm B\kern-.05em{\sc i\kern-.025em b}\kern-.08em
    T\kern-.1667em\lower.7ex\hbox{E}\kern-.125emX}}

\usepackage{svg}
\newcommand{\ie}{{\em i.e.}, }
\newcommand{\eg}{{\em e.g.}, }
\usepackage{lipsum}
\newcommand\blfootnote[1]{%
  \begingroup
  \renewcommand\thefootnote{}\footnote{#1}%
  \addtocounter{footnote}{-1}%
  \endgroup
}

\newcommand{\vspacesections}{\vspace{-0.2mm}}
\newcommand{\mysection}[1]{\vspacesections \section{#1} \vspacesections}

\begin{document}

\title{Architecture-Level Modeling of Photonic Deep Neural Network Accelerators}

\author{
\IEEEauthorblockN{Tanner Andrulis}
\IEEEauthorblockA{\textit{MIT} \\ Cambridge, USA \\ andrulis@mit.edu}
\and
\IEEEauthorblockN{Gohar Irfan Chaudhry}
\IEEEauthorblockA{\textit{MIT} \\ Cambridge, USA \\ girfan@mit.edu}
\and
\IEEEauthorblockN{Vinith M. Suriyakumar}
\IEEEauthorblockA{\textit{MIT} \\ Cambridge, USA \\ vinithms@mit.edu}
\and
\IEEEauthorblockN{Joel S. Emer}
\IEEEauthorblockA{\textit{MIT, Nvidia} \\ Cambridge, USA \\ jsemer@mit.edu}
\and
\IEEEauthorblockN{Vivienne Sze}
\IEEEauthorblockA{\textit{MIT} \\ Cambridge, USA \\ sze@mit.edu}
}

\maketitle

\begin{abstract}
Photonics is a promising technology to accelerate Deep Neural Networks as it can use optical interconnects to reduce data movement energy and it enables low-energy, high-throughput optical-analog computations. 

To realize these benefits in a full system (accelerator + DRAM), designers must ensure that the benefits of using the electrical, optical, analog, and digital domains exceed the costs of converting data between domains. Designers must also consider system-level energy costs such as data fetch from DRAM. Converting data and accessing DRAM can consume significant energy, so to evaluate and explore the photonic system space, there is a need for a tool that can model these full-system considerations.

In this work, we show that similarities between Compute-in-Memory (CiM) and photonics let us use CiM system modeling tools to accurately model photonics systems. Bringing modeling tools to photonics enables evaluation of photonic research in a full-system context, rapid design space exploration, co-design, and comparison between systems.

Using our open-source model, we show that cross-domain conversion and DRAM can consume a significant portion of photonic system energy. We then demonstrate optimizations that reduce conversions and DRAM accesses to improve photonic system energy efficiency by up to $3\times$.

\end{abstract}


\mysection{Introduction} \label{introduction}

% Deep Neural Networks are becoming increasingly complex and the amount of data available for training them is rapidly growing.
% From a computer architecture perspective, this translates into expensive data movement operations that dominate the energy consumption of the overall system.
% To address this, there have been promising advances in using analog photonic architectures for DNN acceleration.
% These systems use several low-energy photonic components that are well-suited for DNN training.
% The inherent parallelism in optical components (from high channel count) provides high bandwidth which enables scaling the systems to perform fixed-point operations.

Deep Neural Networks (DNNs) can be energy-intensive to compute due to the movement of large tensors and the many multiply-accumulate (MAC) operations that they require. To address these challenges, photonic systems (accelerator + DRAM) leverage the digital-electrical ($DE$), analog-electrical ($AE$), digital-optical ($DO$), analog-optical ($AO$) domains. Specifically, optical (\ie $DO$ and $AO$) interconnects can reduce data movement energy while analog (\ie $AE$ and $AO$) computation can reduce MAC energy.
\blfootnote{\label{opensource}The model, tutorials, and examples are available in the CiMLoop~\cite{cimloop} repository at \href{https://github.com/mit-emze/cimloop}{\(\textit{https://github.com/mit-emze/cimloop}\)}.\nocite{timeloop,accelergy,ruby}}

% \textit{Optical computing} leverages low-energy $AO$ interconnects to move data, while \textit{compute-in-memory (CiM)} computes directly in $AE$ or $DE$ memory arrays to avoid moving some DNN tensors. Both approaches can leverage analog computing to reduce MAC energy.

Unfortunately, in a full system, the benefits of these domains can be limited by the costs of other components. Specifically, systems may pay significant energy to convert data between domains~\cite{adept,albireo,lightning,raella,andrulis2024modeling} and to fetch data from DRAM~\cite{albireo,adept}. To evaluate photonic systems, there is a need for a tool that can model these costs. Furthermore, to optimize full systems, the tool must rapidly explore a large co-design space that includes components (\eg data converters, SRAM buffers, optical resonators), architecture (\ie what components are used, how many components, how they connect), workload (\ie DNN layer types, tensor shapes/values), and mapping (\ie how the workload is scheduled onto the architecture). 

Fortunately, these characteristics are not unique to photonics. Analog Compute-in-Memory (CiM) systems have a large full-system co-design space, leverage the advantages of multiple domains ($AE$ and $DE$), and face the challenge of high cross-domain conversion energy. 

In this work, we show that these similarities let us leverage the open-source CiMLoop~\cite{cimloop, accelergy, timeloop, ruby} tool to accurately model photonic systems. Bringing this tool to photonics enables researchers to \textbf{(1)} accurately evaluate and compare research contributions in a full-system context (\eg see how a novel component affects a full system or compare two photonic systems across a range of DNN workloads) \textbf{(2)} perform fast design-space exploration over the large co-design space~\cite{cimloop}, and \textbf{(3)} share knowledge between the photonics and CiM research communities. 

% Unfortunately, it can be a challenge to realize these potential benefits in a full system, in part because energy benefits can be overwhelmed by high-energy off-chip DRAM~\cite{lightning} and by data converters that convert data between domains~\cite{raella}.

% Another challenge is the difficulty in designing dataflows that combine analog/photonic data movement with a traditional digital memory hierarchy, and in ensuring that the data movement energy of such a dataflow does not overwhelm the potential benefits of low-energy computation~\cite{lightning}.

% To address these challenges, there is a need for tools that can model full photonics systems and rapidly explore the photonics design space, but no such tools exist to the authors' knowledge. F

% Unfortunately, while such tools exist for CiM~\cite{cimloop, accelergy, timeloop}, to the authors' knowledge no such tools exist for photonics. 

% Many practical implications arise from the co-existance of analog and digital components in the same system, including components like analog-to-digital (ADC) and digital-to-analog (DAC) converters.
% In addition to these, the impact of data movement between the two domains (e.g., to/from off-chip DRAM) must be considered to accurately evaluate the performance and energy efficiency of these systems.

\mysection{Photonics Modeling Tool}
% \subsection{Component Specifications}
% Our photonic system framework combines a Timeloop-based mapspace explorer and an Acclergy-based energy/area esimtator. We provide implementations of commonly found components in optical systems: Mach-Zender Modulator (MZMs), Microring Resonators (MRRs), O2E (Optical to Electrical converter), E2O (Electrical to Optial converter), and commonly implemented components from CiM such as Analog to Digital Converter (ADC) and the Digital to Analog Converter (DAC). 
The tool takes as input specifications of a DNN workload, components, and architecture as defined in Section~\ref{introduction}. The tool maps the given workload on the architecture and outputs full-system area, energy, and throughput estimations.

\begin{figure}[]
    \centering
    \includegraphics[width=\linewidth, keepaspectratio] {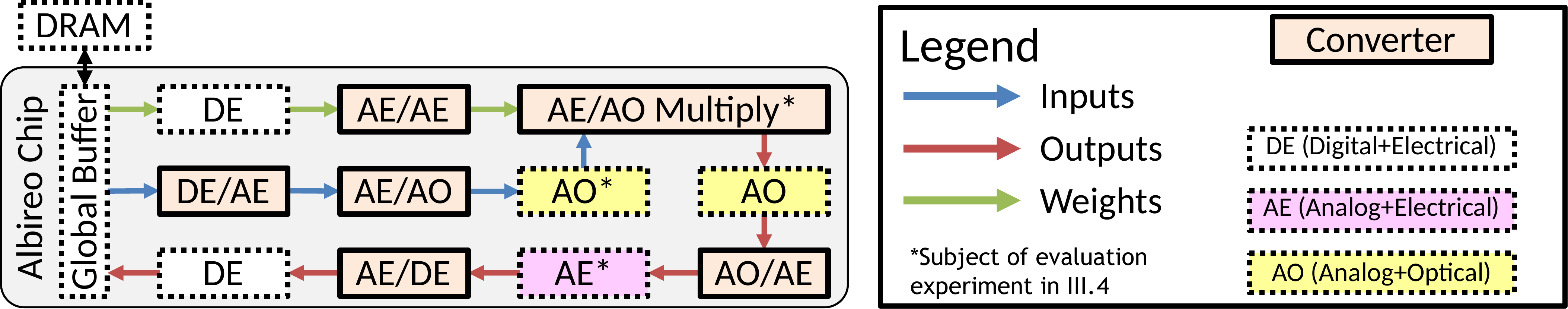}
    \caption{\label{fig:albireo} Albireo architecture. As data traverse the $DE$, $AO$, and $AE$ domains, they leverage different movement and reuse opportunities but pay energy for data converters, notated $X/Y$ for conversion from domain $X$ to domain $Y$.}
\end{figure}

In this work, we model the Albireo~\cite{albireo} photonic system, which leverages the $DE$, $AE$, and $AO$ domains ($DO$ is used in~\cite{jouppi2023tpu}). Fig.~\ref{fig:albireo} shows how Albireo moves data through each domain. $DE$ can reuse data spatially and temporally with multicast/reduction networks, buffer hierarchies, and DRAM. $AE$ can use low-energy analog multiplications, additions, and data movement. $AO$ can further reduce data movement energy with low-energy, high-throughput optical interconnects.

% \subsubsection{Component Specifications}
% We model Albireo's components, listed below in \textbf{bold}. Albireo's components are commonly used in other photonic systems~\cite{hamerly_edge_2021,doi:10.1126/science.abq8271,shen_deep_2017,pixel,holylight,pcnna,DNNARA,feldmann_parallel_2021,adept,Bandyopadhyay2022SingleCP}, and these models can easily be used in other user-defined architectures.

% \textbf{Lasers} act as a source of light. \textbf{Mach-Zender Modulators (MZMs)} attenuate light based on an analog signal, and can act as analog-photonic converters or analog-photonic multipliers. \textbf{Microring Resonators (MRRs)}, \textbf{arrayed waveguide grating}, and \textbf{star couplers} act as switches, demultiplexers, and any-to-any crossbars, respectively, for light on chip. Finally, \textbf{photodiodes} with \textbf{transimpediance amplifiers (TIAs)} act as photonic-to-analog converters. \textbf{Digital-analog-converters (DACs)} and \textbf{analog-digital-converters (ADCs)} convert values to/from the analog domain, while \textbf{digital buffers} store and reuse data.

A key design decision in photonic systems is where to cross between domains. This is because \textbf{(1)} energy and area costs of data movement, data reuse, and computations change significantly between domains~\cite{cleocourse} and \textbf{(2)} crossing domains requires high-energy data converters, shown in Fig.~\ref{fig:albireo} as $X/Y$ for a domain crossing from $X$ to $Y$. In particular, $AE/DE$ and $DE/AE$, commonly known as analog-to-digital and digital-to-analog converters, can consume significant energy~\cite{raella,andrulis2024modeling}.

Reducing data converter energy is a key challenge in both CiM and photonic systems~\cite{raella, lightning}. The number of conversions, and thus data conversion energy, can be reduced by leveraging \textit{reuse}: converting a value into a domain and reusing the converted value multiple times in that domain (\eg convert $DE$ value $V_{DE}$ to $AE$ value $V_{AE}$ with a $DE/AE$ converter, then use the $V_{AE}$ for multiple $AE$-domain computations)~\cite{raella}.

% The decision of where to put domain crossings affects both the number and energy of domain crossings and the energy of other operations. Fig.~\ref{fig:domains} shows an example, where we compare two locations for a domain crossing. The first option converts X once from domains A to B, reusing the converted value to compute $f(X)$ and $g(X)$ in domain B. The second computes $f(X)$ and $g(X)$ in domain A then converts each result to domain B, incurring two conversions. Not only does this decision affect the number of conversions, but it also affects the energy of conversions~\cite{}

% \textit{Many on-chip actions, including , the number and energy of conversions,}

% \begin{figure}[]
%     \centering
%     \includegraphics[width=\linewidth, keepaspectratio] {images/domains_svg-raw.pdf}
%     \caption{\label{fig:domains} Albireo architecture. As data traverse the $DE$, $AO$, and $AE$ domains, they leverage different movement and reuse opportunities but pay energy for data converters, notated $X/Y$ for a conversion from domain $X$ to domain $Y$.}
% \end{figure}

%   To reduce data conversion energy and area, systems can reduce the number of conversions by spatially and temporally reusing converted values in the $AE$ and $AO$ domains. Furthermore, $AO$ can reuse data across multiple wavelengths of light (known as wavelength multiplexing).

To model Albireo, we augment CiMLoop's $DE$ and $AE$ component library with $AO$ components such as microring resonators, star couplers, lasers, Mach-Zender modulators, and photodiodes~\cite{hamerly_edge_2021,doi:10.1126/science.abq8271,shen_deep_2017,pixel,holylight,pcnna,DNNARA,feldmann_parallel_2021,adept,Bandyopadhyay2022SingleCP}. We expose each component's data movement and reuse opportunities to CiMLoop's mapper, which finds mappings that leverage available reuse to minimize energy-intensive conversions and DRAM accesses.

\mysection{Evaluation}
We first validate the energy accuracy of the modeling tool. We then evaluate throughput on two DNN workloads, evaluate the full Albireo system with DRAM, and explore architectural approaches to reducing data converter energy.

\subsubsection{Accelerator Energy Breakdown} 
\begin{figure}[]
    \centering
    \includegraphics[width=\linewidth, keepaspectratio] {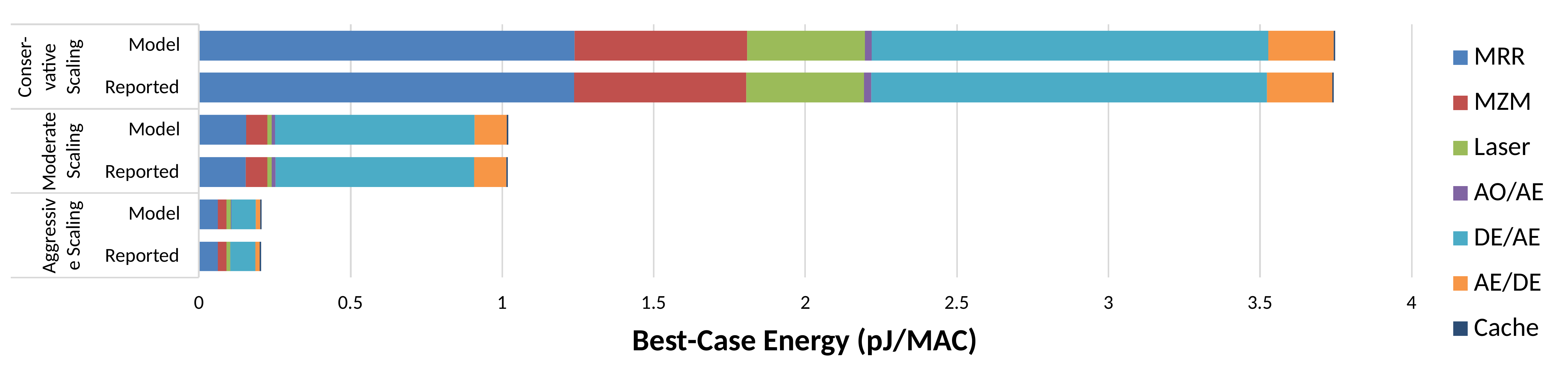}
    \caption{\label{fig:energy_breakdown} Energy breakdown validation.}
\end{figure}

Fig.~\ref{fig:energy_breakdown} shows the modeled versus reported energy breakdown results. We show all components in the Albireo~\cite{albireo} paper, including the accelerator and an off-chip laser. The average overall energy error is $0.4\%$.

\subsubsection{Throughput} 
\begin{figure}[]
    \centering
    \includegraphics[width=\linewidth, keepaspectratio] {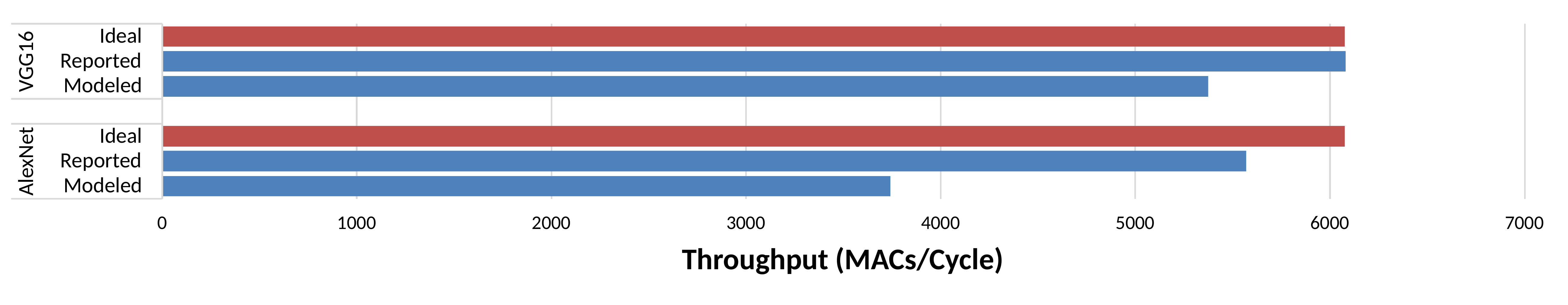}
    \caption{\label{fig:throughput}Throughput for two DNN workloads. CiMLoop captures underutilization, which significantly degraded throughput for Albireo running AlexNet.}
\end{figure}

Fig.~\ref{fig:throughput} shows the modeled versus reported throughput for VGG16~\cite{vgg} and AlexNet~\cite{alexnet}. We also include an ideal throughput, which assumes 100\% compute unit utilization. While results in~\cite{albireo} are near ideal, we find that modeled throughput is significantly lower when accounting for underutilization due to different DNN weight tensor shapes. In particular, Albireo is designed for unstrided convolutional layers, and fully-connected and strided convolutional severely underutilize Albireo's compute units. These differences illustrate the importance of using a model that can accurately evaluate throughput~\cite{cimloop} by capturing many sources of underutilization.

\subsubsection{Full-System (Accelerator+DRAM)} Albireo fetches operands from DRAM, but the paper omits DRAM energy. We connect Albireo to DRAM in our model to see how it impacts energy. We evaluate two configurations from the Albireo paper based on aggressive (high-energy) and conservative (low-energy) scaling projections for future optical components.

\begin{figure}[]
    \centering
    \includegraphics[width=\linewidth, keepaspectratio] {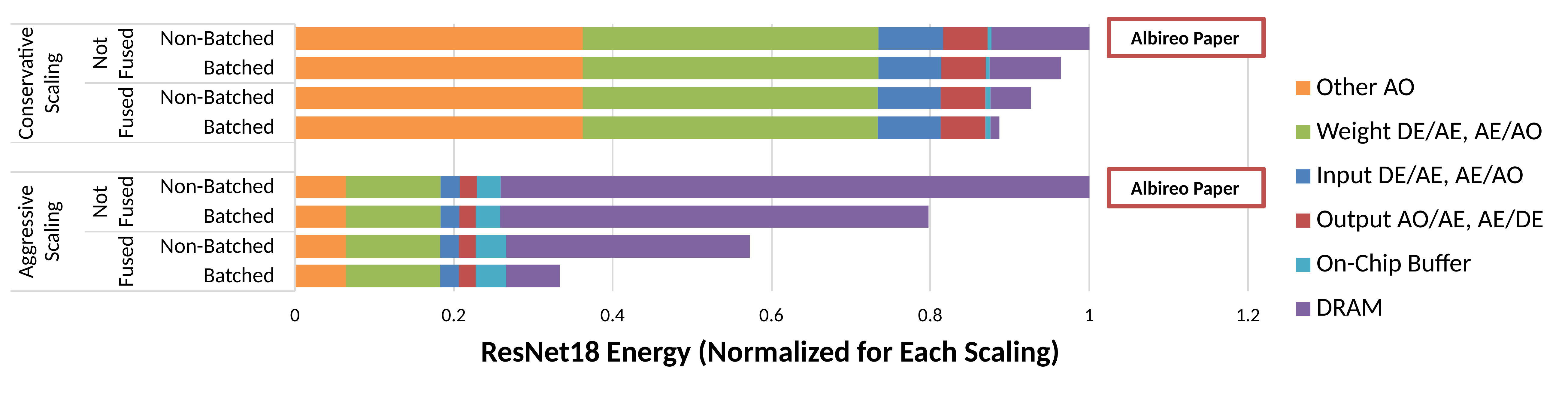}
    \caption{\label{fig:memory_exploration} Memory exploration. Aggressively-scaled Albireo is dominated by DRAM energy. DRAM-energy-reducing operations such as batching and fusion are required to realize the benefits of aggressive scaling, }
\end{figure}

Fig.~\ref{fig:memory_exploration} shows that for the conservatively-scaled Albireo, DRAM consumes little overall energy, but for the aggressively-scaled Albireo, DRAM consumes 75\% of overall system energy. To achieve the potential energy benefits of aggressive scaling, it is critical to reduce DRAM energy. We explore two strategies to do so. First, we batch inputs and outputs to amortize weight movement energy. Next, we keep inputs and outputs on-chip in the global buffer between layers rather than fetching them from DRAM~\cite{looptree}. The former strategy increases latency, while the latter requires a larger global buffer and therefore more global buffer energy. Using both of these strategies together, we can reduce aggressively-scaled system energy by 67\% ($3\times$ improvement).

% This test illustrates first that given a published architecture CiMLoop can in a system with a memory hierarchy, which is important because the memory hierarchy can significantly impact overall energy. Second, it shows that CiMLoop can be used to explore 

\subsubsection{Architecture Exploration} 
\begin{figure}[]
    \centering
    \includegraphics[width=\linewidth, keepaspectratio] {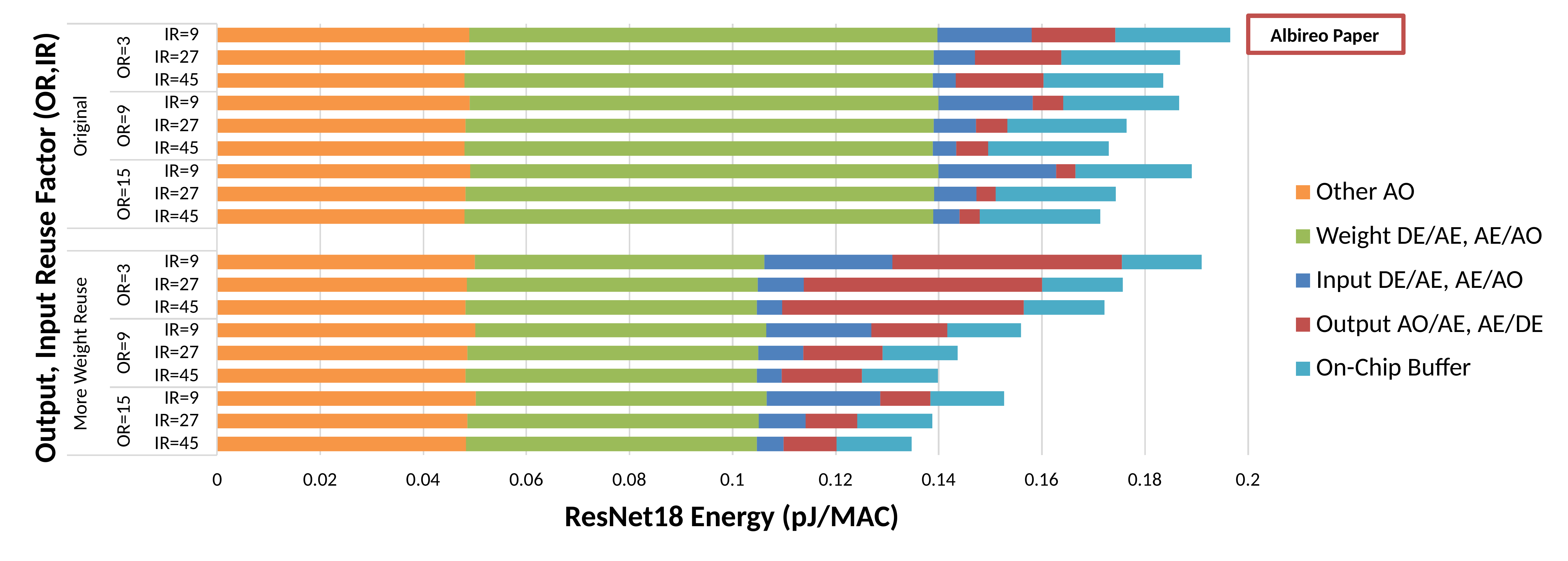}
    \caption{\label{fig:architecture_exploration}Increasing the amount of reuse in the analog and photonic domains can reduce data conversion energy, leading to a lower-energy system.}
\end{figure}

Albireo pays significant energy for data converters. To decrease data conversion energy, we can reduce the number of conversions by converting a value once and reusing the converted value spatially among multiple components~\cite{raella}. In Fig.~\ref{fig:architecture_exploration}, we explore variations of the aggressively-scaled Albireo architecture that modify the amount of data reuse. Note that increasing $AO$ reuse will also decrease $DE/AE$ and $AE/DE$ energy because Albireo uses $AE$ as an intermediate between $DE$ and $AO$.

We modify the $AE/AO\ Multiply^*$ block in Fig.~\ref{fig:albireo}, connecting more $AO$ components to spatially reuse $AE$ weights (lower weight conversion energy) rather than reusing $AO$ inputs and outputs (higher input and output conversion energy). To reduce input conversion energy, we increase the number of components that spatially reuse $AO$ inputs in the $AO^*$ block. To reduce output conversion energy, we increase the number of output-reusing $AE$ components by modifying the $AE^*$ block. We find that increasing reuse can reduce data converter energy by 42\% and can reduce accelerator energy by 31\%.

\bibliographystyle{IEEEtran}
\bibliography{main.bib}

\end{document}